\documentclass[12pt]{article}
\usepackage{amssymb,amsmath,esint,titlesec}
\titleformat*{\section}{\normalsize\bf}
\titleformat*{\subsection}{\small\bf}

\begin{document}


\begin{titlepage}

\setlength{\baselineskip}{18pt}

                               \vspace*{0mm}

                             \begin{center}

{\LARGE\bf Non-linear Fokker-Planck equations from \\

                                               \vspace{5mm}                                           
 
                       conformal metrics and scalar curvature}

                                   \vspace{40mm}

              \Large\sf  NIKOLAOS \   KALOGEROPOULOS  $^\dagger$\\

                            \vspace{1mm}

  \normalsize\sf  Department of Mathematics and Natural Sciences,\\
                           The American University of Iraq, Sulaimani,\\
                           Kirkuk Main Road, Sulaimani,\\
                              Kurdistan Region, Iraq. \\

                            \vspace{2mm}
                         
                                    \end{center}

                            \vspace{20mm}

                     \centerline{\normalsize\bf Abstract}
                     
                           \vspace{1mm}
                     
\normalsize\rm\setlength{\baselineskip}{18pt} 

We present an argument which intends to explore a potential geometric origin of a class of non-linear Fokker-Planck equations 
related to the mesoscopic behavior of systems conjecturally described by the $q$-entropy.
We argue that the appearance of the non-linear term(s) in such equations can be ascribed to the fact that the 
effective mesoscopic metric describing the behavior of the underlying system may not be the originally chosen one, but a conformal deformation of it. 
Motivated by Liouville's theorem, we highlight the role played by the scalar curvature of conformally related metrics 
in establishing such a non-linear Fokker-Planck equation. \\

                           \vfill

                                                 \hspace*{19.5mm} Scalar curvature.\\
                                                                         
                             \vfill

\noindent\rule{9cm}{0.2mm}\\  
   \noindent   $^\dagger$ {\footnotesize\rm Electronic mail: \ \  \  \normalsize{nikos.physikos@gmail.com}}\\

\end{titlepage}
 

                                                                                \newpage                 

\rm\normalsize
\setlength{\baselineskip}{18pt}

\section{Introduction}

The linear Fokker-Planck equation is an important  mesoscopic equation, which is extensively used in various parts of equilibrium and non-equilibrium Statistical Mechanics and their applications \cite{vanKampen, Risken}. It can be derived from a Kramers-Moyal expansion of the master equation subject to some reasonable approximations 
\cite{vanKampen, Risken}. Its mathematical properties and those of its solutions  occupy a distinguished position in the theory of stochastic processes. 
It can be seen as the prototypical example of a mesoscopic, semi-empirically/phenomenologically derived, 
evolution equation of the probability distribution function describing the statistical behavior of the system under study. \\

For various reasons, a need has arisen during the last decades in formulating non-linear counterparts of the linear Fokker-Planck equation. 
Such non-linear equations \cite{Frank} are usually derived using additional assumptions or auxiliary constitutive relations specific to the class of systems under consideration. 
Our motivation for the present work, is the considerable amount of  effort which has been invested in establishing connections between solutions of 
non-linear Fokker-Planck equations  and extremizing distributions, known as $q$-exponentials \cite{Tsallis-book}, of variational problems involving the $q$-entropy 
\cite{Tsallis-book, PP, TB, Borland, BPPP,  MPP, Shiino, Beck, MMPL, Frank, CN1, Chav, NCR, SCN1, SNT, FC, SCN2, ASMNC, Shiino2, RNC, LB, ABT, RCN, SANC, PCNT, CN2}.\\

In the present work, we explore the possibility of using the underlying geometry of the mesoscopic (coarse-grained) system in order to set up  non-linear Fokker-Planck equations. 
We rely on some of our prior work pertinent to the behavior of systems conjeturally described by the $q$-entropy such as the metric and measure properties of 
the underlying system under (quasi-)conformal maps as conjectured in \cite{NK1, NK2}. Our work  has commonalities with many works in the existing literature,  
but it is different in that it relies heavily on a geometric treatment of the effective dynamics. We work in a Riemannian context and use the far more amenable to analysis conformal 
transformations of the mesoscopic metric to construct Fokker-Planck equations whose solutions contain ansatze similar or reminiscent of  the $q$-exponentials. The latter, as is well-known, 
 are functions extremizing the $q$-entropy functionals under appropriate constraints. We notice the important role that the scalar curvature of the mesoscopic 
metric plays in this metric approach. \\

In Section 2, we provide some well-known facts about the phase space of mechanical systems and the role of metric deformations of the underlying Riemannian metric 
under conformal deformations, as a motivation for the subsequent discussion.  In Section 3, we discuss the behavior of the scalar curvature under conformal deformations of the metric 
and provide some ideas for setting up corresponding non-linear Fokker-Planck equations.  Section 4 contains some conclusions and points out toward a general 
research direction in the future.    \\        


\section{Linear and non-linear Fokker-Planck equations}

\subsection{A few comments on Mechanics}
Throughout this work, we assume that have  a particle system of $N$ degrees of freedom, which as is well-known, has a 
phase space \ $\mathfrak{M}$ \ of finite dimension \ $2N$. \ A chart/local 
coordinate system  of \ $\mathfrak{M}$ \ is parametrized as \ 
$(x^1,\ldots,x^n,x^{N+1},\ldots,x^{2N})$. \  The phase space is a symplectic manifold, endowed with the standard symplectic form
\begin{equation} 
\omega \ = \  \sum_{i=1}^N \ dx^i \wedge dx^{i+N} 
\end{equation}
where \ $d$  \ stands for the exterior derivative and \ $\wedge$ \ for the wedge product of forms. 
Any symplectic manifold \cite{McDS}   possesses a contractible set of compatible almost complex structures. Compatibility signifies two things \cite{McDS}: 
the tameness condition of the almost complex structure \ $\mathbb{J}$ \ by $\omega$, namely 
\begin{equation}
           \omega(X, \mathbb{J}X) > 0,   
\end{equation}
and the invariance of the symplectic structure \ $\omega$ \ under the action of the almost complex structure\ $\mathbb{J}$, \  namely
\begin{equation}
                 \omega(\mathbb{J}X, \mathbb{J}Y) = \omega(X,Y)
\end{equation}
where \ $X,\ Y$ \  are tangent vector fields of \ $\mathfrak{M}$. \
For any such compatible almost complex structure \ $\mathbb{J}$, \  a Riemannian metric \ $\mathfrak{g}$ \ can be constructed on \ $\mathfrak{M}$ \ as \cite{McDS}
\begin{equation} 
               \mathfrak{g}(X,Y) = \omega(X,\mathbb{J}Y)
\end{equation}
In a Hamiltonian system, one has a Hamiltonian \ 
$\mathcal{H}$, \  whose associated Hamiltonian vector field \ $X_\mathcal{H}$ \ is given by
\begin{equation}   
              \omega (X_\mathcal{H}, \cdot) = - d\mathcal{H}
\end{equation}
 The evolution of the system is given as a curve in \ $\mathfrak{M}$ \ whose tangent at 
every point is \ $X_\mathcal{H}$. \  If one has an isolated system, then it is described by an autonomous Hamltonian, and the geodesics of \ $\mathfrak{g}$ \ are the integral 
curves of \  $X_\mathcal{H}$, \  which provide the evolution of the system. Finding explicitly such geodesics is a practically intractable problem \cite{Berger}, 
even for systems having few 
degrees of freedom, let alone when \ $N \sim N_A$, \ where \ $N_A$ \ is Avogradro's number.\\


\subsection{From Mechanics to Statistical Mechanics}

 For such a large number degrees of freedom, one does not need to know
the behavior of one particular geodesic, but rather pays attention to the behavior of a set of geodesics in the phase space \ $\mathfrak{M}$. \ 
Such geodesics arise from using the same Hamiltonian \ $\mathcal{H}$ \ but having
different  initial conditions. This is the ensemble idea due to J.W. Gibbs. 
By hindsight, using such  a set of initial conditions is a necessity. It is not only the underlying quantum nature of all physical systems which precludes 
knowing their exact state with infinite precision, but also the conclusion of  the symplectic non-squeezing theorem \cite{Gromov1}, and the related development
of symplectic capacities \cite{McDS} which puts constraints, in addition to the volume preservation expressed by the Liouville' theorem, to such a phase space evolution.\\   

Since we are interested only in mesoscopic properties of the underlying system, some form of coarse-graining of any microscopic quantity is usually performed. This is a rather subtle 
matter, not toally devoid of controversy even today.  We do have to point out the interesting proposal of \cite{deGosson1, deGosson2} about using 
``quantum blobs" in the linear symplectic category to perform such a coarse-graining. We looked on coarse-graining from the viewpoint of  convexity in \cite{NK3}. \\

At the level of the underlying metric \ $\mathfrak{g}$, \ the process of considering sets of initial conditions evolving under the action of the Hamiltonian
amounts to actually calculating the Ricci and the scalar curvatures, at least in a purely Riemannian context.  
After such a statistical averaging is performed, one focuses on the resulting mesoscopic description. Therefore one is content with calculated averages over the microscopic 
quantities of interest taken over the set of geodesics of the phase space \  $\mathfrak{M}$  \ as these are the only features that matter in the description of the system at a mesoscopic scale.  
This is an effective description involving the omission of the contributions of, possibly quite a few and in an appropriate thermodynamic limit of even infinitely  many degrees of freedom. 
As a result of such a coarse-graining process, one gets an integrable, of even smooth,  
probability distribution function \ $\rho$ \  which determines the evolution of the mesoscopic quantities of the system. \\

 As is well-known, Liouville's theorem \cite{McDS} provides a constraint for
the Hamiltonian evolution of the underlying (microcscopic) dynamical system. It states that the symplectic volume is conserved under a Hamiltonian flow.  The exact formulation is that 
\begin{equation}     
                                   \mathcal{L}_{X_\mathcal{H}} \omega^N \ = \ 0
\end{equation}
where \ $\mathcal{L}$ \ stands for the Lie derivative along its subscript, the Hamiltonian vector field \ $X_\mathcal{H}$. \ 
In an autonomous system, the Hamiltonian evolution takes place on a constant  energy hypersurface of the phase space \ $\mathfrak{M}$ \ to which it provides its 1-dimensional characteristic foliation, implemented  by the set of integral curves of the vector field \ $X_\mathcal{H}$ \cite{McDS}.
\ Passing to the mesoscopic level and assuming that the effective Hamiltonian of the system exists and is still autonomous,
this corresponds to a preservation of the probability distribution \ $\rho$ \ along the effective Hamiltonian flow on \ $\mathcal{M}$. \  Such a probability distribution \ $\rho$ \ is a 
microcanonical distribution. The dynamical behavior of \ $\rho$ \ on the constant energy hypersurface of \ $\mathcal{M}$ \ on which the mesoscopic 
system evolves is also provided by an effective Liouville equation, which when expressed in terms of \ $\rho$ \ states that \ $\rho$ \ should be constant, namely that 
\begin{equation}          
          \frac{d\rho}{dt} \ = \   \frac{\partial\rho}{\partial t} + \{ \rho, H \} = 0  
\end{equation}
where the curly brackets in (7) are the Poisson brackets,  defined for infinitely smooth functions \  $f_1, f_2 \in C^\infty (\mathcal{M})$, \ by  
\begin{equation}
            \{ f_1, f_2 \} \ = \ \omega(X_1, X_2)
\end{equation}
where the relation between the functions \ $f_1, \ f_2$ \ and the corresponding vector fields \ $X_1, \ X_2$ \ respectively, is given by (5).\\ 

At this stage, it is not clear at all what may the significance be of the underlying metric \ $\mathfrak{g}$ \ of the phase space \ $\mathfrak{M}$. \ 
Most treatments tend to assume that the mesoscopic behavior of the system is provided by a metric \ $\mathbf{g}$ \  of the same form as the Riemannian metric 
\  $\mathfrak{g}$ \ that was initially chosen for \ $\mathfrak{M}$ \  through (4).  
But this similarity of functional forms between \ $\mathfrak{g}$ \  and \ $\mathbf{g}$ \ may not necessarily be a correct, or an  effective way for encoding the 
pertinent physical properties of the system under study.  It is entirely possible, that due to averaging and coarse-graining 
in the transition from the microscopic to the mesoscopic system, that \ $\mathbf{g}$ \  
on the mesoscopic effective phase space \ $\mathcal{M}$ \  should be  different from the induced metric by \ $(\mathfrak{M}, \mathfrak{g})$.   \\   


\subsection{On Fokker-Planck equations}

 The linear Fokker-Planck equation describes the mesoscopic evolution of the afore-mentioned probability distribution \ $\rho$ \cite{vanKampen}.  
The probabilistic nature of \ $\rho$ \ arises because we have chosen to ignore many features of the underlying system described by \ $\mathcal{H}$ \ which should be, one assumes, statistically insignificant for describing the physical properties and calculating the physical quantities of interest. 
The combined effect of the reduction of dynamics due to averaging and coarse-graining,  gives rise to a collective effect which is expressed as ``noise" 
in the corresponding Langevin approach \cite{vanKampen}. 
As a result of such necessary reductions and approximations, the phase space of the underlying ``microscopic" system \  $\mathfrak{M}$ \ is getting replaced by the 
effective phase space \ $\mathcal{M}$ \ of the effective/mesoscopic model on which the probability distribution \ $\rho$ \ is defined. The dimension of \ $\mathcal{M}$, 
\ call it \ $n$, \ is the same or lower than that of \ $\mathfrak{M}$ \ in such a description. \\

Let \ $(z_1, \ldots, z_n)$ \ be local coordinates of \ $\mathcal{M}$ \ and  that \ $n \leq N$. \ Hence, at the mesoscopic level, the 
effective phase space is a Riemannian manifold \ $\mathcal{M}$ \ endowed with an effective metric \ $\mathbf{g}$. \ The Fokker-Planck equation describes the evolution of 
\ $\rho (t, z_1,\ldots, z_n)$ \ on such a background. The linear Fokker-Planck equation describing the evolution of such \ $\rho (t, z_1, \ldots, z_n)$, \ 
assuming an isotropic diffusion coefficient \ $D$ \ on \ $\mathcal{M}$ \ for simplicity,  is given by 
\begin{equation}
      \frac{\partial \rho(t, z_1,\ldots, z_n)}{\partial t} \ = \ D \ \nabla^2 \rho(t, z_1, \ldots, z_n)
\end{equation} 
where \ $\nabla$ \ indicates the gradient, and \ $\nabla^2$ \ the Laplacian of \  $(\mathcal{M}, \mathbf{g})$. \ 
In (9), \ $t\in[0,+\infty)$ \  is an evolution parameter. It does not have to be the physical time, 
even though on many occasions it is. We have set the drift term which appears in the derivation of  the Fokker-Planck equation 
(9) equal to zero for simplicity, at least initially, as it will naturally emerge again in the subsequent arguments.  
The diffusion coefficient \ $D$ \  is assumed to be constant in (9). If it were not, then (9) would be modified to read 
\begin{equation}
        \frac{\partial \rho(t, z_1, \ldots, z_n)}{\partial t} \ = \ \nabla (D(z_1,\ldots,z_n) \nabla \rho(t, z_1, \ldots, z_n))
\end{equation}
From now on, we will restrict our attention in the rest of this work only to the cases of Hamiltonian systems of many degrees of freedom whose statistical behavior is
conjecturally described by the Tsallis/$q$- (henceforth $q$-) entropy \cite{Tsallis-book, Tsallis}. This is a single-parameter family of entropic functions
which is defined for discrete sets of outcomes parametrized by an index set \ $I$ \ as
\begin{equation} 
          \mathcal{S}_q[\{ p_i \}] \ = \  k_B \ \frac{1}{q-1} \left( 1-\sum_{i\in I} p_i^q \right) 
\end{equation}
The obvious, but not necessarily correct,  continuum generalization of this function, for a probability distribution \ $\rho$ \ on some sample 
space \ $\Omega$ \ having Riemannian volume \ $\mathrm{vol}$, \  is  the functional 
\begin{equation}
           \mathcal{S}_q [\rho ] \ = \ k_B \  \frac{1}{q-1} \left\{ 1- \int_\Omega [\rho(x)]^q \ \mathrm{dvol}(x)  \right\}
\end{equation}
where Boltzmann's constant \  $k_B$ \ will be set equal to unit for brevity, and where the  non-additive/entropic  parameter \ $q$ \ will be assumed to take values in 
the set of reals \ $\mathbb{R}$. \\
 
There is a wide-spread suspicion/conjecture that systems whose theormodynamic behavior is captured by the $q$-entropy, are described mesoscopically 
by non-Markovian processes \cite{Tsallis-book}, for which the Smoluchovskii/forward Kolmogorov  equation, which is used to derive the Fokker-Planck equation (9), 
does not hold \cite{vanKampen}. 
Hence, if one wishes to keep an effective mesoscopic/kinetic treatment of the statistical behavior of the system, 
one should appropriately modify (9) to also cover the cases of non-Markovian processes. 
 One possible way to do this has been to replace (9), in a largely empirical/ad hoc manner, by a non-linear counterpart. 
Probably the simplest non-linear analogue of (9) is the well-known porous medium equation \cite{Vazquez}
\begin{equation}
          \frac{\partial \rho(t, z_1, \ldots, z_n)}{\partial t} \ = \ D \ \nabla^2 (\left[\rho (t,z_1, \ldots, z_n)\right]^m) 
\end{equation}   
 where \ $m\in \mathbb{R}$ \  is a phenomenological constant  usually introduced by hand to match either obtained data or desirable properties of some particular 
ansatze serving as proposed solutions.  \\

A question that may be worth pursuing at this point, is how to ``derive"  (13), if not from first principles as would be ideal, at least by using some 
``reasonable" assumptions at the mesoscopic level. We would like, ideally,  to avoid as much as possible making completely ad hoc or purely phenomenological assumptions 
on the path leading to (13). 
One way to achieve this is to allow for variations of the diffusion coefficient in (9), (10) that depend 
on the points of \ $\mathcal{M}$ \ explicitly, but also through the probability distribution \ $\rho$ \  itself. 
By that, we mean an equation of the form 
\begin{equation}
    \frac{\partial \rho (t, z_1, \ldots z_n)}{\partial t} \ = \ \nabla [D(z_1, \ldots z_n, \rho(t, z_1, \ldots z_n)) \ \nabla \rho(t, z_1,\ldots, z_n)]
\end{equation}    
The question of interest is now reduced into how such a probability-dependent diffusion coefficient 
\begin{equation}
D(z_1, \ldots z_n, \rho (t, z_1, \ldots, z_n))
\end{equation}
may arise.
Our proposal is that such a behavior may be ascribed to the conformal covariance of the mesoscopic quantities used for the system, 
whose thermodynamic behavior is described by the $q$-entropy. \\

We pointed out above that there is no a priori reason why the underlying metric \ $\mathfrak{g}$ \ of the phase space \ $\mathfrak{M}$, \ and the metric \ $\mathbf{g}$ \ 
of the mesoscopic/effective phase space \ $\mathfrak{M}$ \ should be the same, or even related in any obvious way. Coarse-graining is a non-trivial process whose foundations 
need to be further elucidated, at the current level of understanding of the theory. 
We believe that one part of a coarse-graining process in classical Mechanics has to do with the conclusion of the symplectic 
non-squeezing theorem \cite{McDS, Gromov1, deGosson1, deGosson2} which puts constraints on the behavior of Hamiltonian maps and flows that go beyond those of phase space volume preservation expressed  by Liouville's theorem. The direct  implications for Statistical Mechanics, if any, of the symplectic non-squeezing theorem are currently unknown.    \\
  

\subsection{On the role of conformal transformations}
   
The two metrics \ $\mathfrak{g}$ \ and \ $\mathbf{g}$ \ are usually taken to have exactly the same form,
based on general assumptions of simplicity. But such requirement of these two metrics having the same form, simple and ``reasonable" even though unjustified, 
may be too restrictive. We tend to suspect that conformal transformations may play a 
fundamental role in the underlying mesoscopic dynamics, for at least the following four  reasons.  These provide some form of circumstantial evidence toward the role of conformal 
transformations  in relating the metrics \ $\mathfrak{g}$ \ and \ $\mathbf{g}$. \ 
We have to stress again that our hand-waving arguments are applicable only to cases of systems described by $q$-entropy. But  even for such systems, we cannot assume that  our treatment 
would be applicable to  all such systems. Obviously then, our arguments  are not meant to be applicable to general Hamiltonian systems of many degrees of freedom.     \\ 

First, the most naive statement goes as follows: Let us for a moment forget about the possible existence of the Riemannian metric (4).
It is well-known that symplectic manifolds such as \ $\mathfrak{M}$ \ lack local structure. Indeed, according to 
Darboux's theorem \cite{McDS} all symplectic forms on a symplectic manifold are locally diffeomorphic to (1). 
One immediately notices that  under a dilatation 
\begin{equation}
           z_i \ \longmapsto \ \lambda z_i,   \hspace{5mm}   i=1,\ldots, n,  \hspace{5mm} \lambda\in\mathbb{R}
\end{equation}
the symplectic form (1) transforms covariantly, namely
\begin{equation}
                  \omega \  \longmapsto \  \lambda^2\omega       
\end{equation}
So all the local structure of a symplectic manifold, like \ $\mathfrak{M}$ \ essentially remains invariant under rigid rescalings. This is only locally true on \ $\mathfrak{M}$, \ 
because  globally \ $\mathfrak{M}$ \ can be considerably different from \ $\mathbb{R}^{2N}$. \ So, at a very fundamental level one can interpret the local scaling symmetry 
of the symplectic form of \ $\mathfrak{M}$ \  as a ``raw" property, which however can be, and usually is, violated by derived quantities of interest once the presence  of 
the Riemannian metric (4) is taken into account.
Therefore, assuming a conformal symmetry for the effective metric \ $\mathbf{g}$ \ in the mesoscopic description of the system, is like going back to the scaling symmetry of 
the underlying phase space \ $\mathfrak{M}$, \ but after having travelled through the circuitous route of the path/geodesic space of \ $\mathfrak{M}$ \ and its subsequent 
coarse graining.\\        
      
Second, from the earliest days of the $q$-entropy, scale invariance has played a significant role in its development \cite{Tsallis-book}. Fractals, which have been the motivation 
for the introduction of $q$-entropy, by Tsallis in \cite{Tsallis} in the Physics community, many times exhibit scale invariance \cite{Falconer} when they are self-similar. 
Such fractals can be frequently seen to be the result of iterated function systems. Conformal invariance is a strengthening of this scale invariance.\\

A third, and closely related to the second, reason is that  as has been noticed the $q$-entropy's  underlying functional form (11), (12) can be seen as the outcome of the application 
of the Jackson derivative
\begin{equation} 
        d_q f(x) = \frac{f(qx) - f(x)}{qx-x}
\end{equation}
 as \cite{Abe}
\begin{equation}
     \mathcal{S}_q[\{ p_i \}] = \left.  - d_q \left( \sum_{i\in I} p_i^x\right)\right|_{x=1}
\end{equation}   
The Jackson derivative  suggests the important role that dilatations  play for the systems described by the $q$-entropy, at least at the mesoscopic level. 
A simple behavior under dilatations is an indication of a potential underlying scale covariance. 
Covariance under conformal maps is a stronger assumption. But it is  desirable on theoretical grounds, in order to get a 
better control of some general common features of systems described by the $q$-entropy. \\

Fourth, such a scale covariance is also seen in solutions of the porous medium equation (13). Let \ $\rho$ \ be one solution of (13). 
Then we can easily verify that the one-parameter set of functions  
\begin{equation}  
                 \rho_\lambda (t, z_1, \ldots, z_n)  \ = \ \lambda^\alpha \rho(\lambda t, \lambda^\beta z_1, \ldots, \lambda^\beta z_n)
\end{equation}
with 
\begin{equation}
                \alpha (m-1) +2\beta \ = \ 0, \hspace{15mm} \alpha > 0, \ \beta > 0 
\end{equation}
for any \ $\lambda >0$ \ also satisfies (13) \cite{Vazquez}.  Hence one can intepret the one-parameter family of scaling transformations (20) as mapping one solution of (13) to another. 
This motivates someone to search for scale invariant solutions of the form
\begin{equation}
              \rho (t, z_1, \ldots, z_n) \ = \ \lambda^{-\alpha} \rho(1, \lambda^{-\beta}z_1, \ldots, \lambda^{-\beta}z_n) \ \equiv 
                                                              \ \lambda^{-\alpha}\ \widetilde{\rho}(\lambda^{-\beta}z_1, \ldots, \lambda^{-\beta}z_n) 
\end{equation} 
Assuming, moreover, that the function \ $\widetilde{\rho}$ \  is radial, meaning that 
\begin{equation}
           \widetilde{\rho} (t, z_1,\ldots, z_n) \ = \ \widetilde{\rho}(\lVert z \rVert)
\end{equation}
one eventually reaches the Barenblatt/ZKB family of solutions \cite{Vazquez}
\begin{equation}
              w_m (t, z_1, \ldots, z_n) \ = \ t^{-\alpha} \left\{ \left( c-\frac{\beta (m-1)}{2m} \frac{\lVert z \rVert^2}{t^{2\beta}} \right)_+  \right\}^\frac{1}{m-1}  
\end{equation}
of the porous medium equation (13). In (24), \ $c>0$ \  is a constant, $\lVert z \rVert$ \ indicates the Euclidean norm of \ $(z_1, \ldots, z_n) \in\mathbb{R}^n$, \ and 
\begin{equation}
                        \alpha \ = \ \frac{n}{n(m-1)+2}, \hspace{15mm} \beta = \frac{\alpha}{n}
\end{equation}
One notices the formal similarity between (24) and the $q$-exponentials \cite{Tsallis-book}
\begin{equation}
                        e_q(x) \ = \ \left(1+(1-q)x\right)^\frac{1}{1-q} _+
\end{equation}
where 
\begin{equation}
        (x)_+ \ = \ \max\left\{0,x \right\}
\end{equation}
in (24),(26). The q-exponentials (26)  are the extremizing distributions of the $q$-entropy (11), (12) under micro-canonical or canonical constraints. Due to this similarity, one 
suspects that an underlying scale invariance on the effective phase space \ $\mathcal{M}$, \ strengthened to conformal invariance in order to be make things more manageable, 
may result in  Fokker-Planck equations that may be of interest for the mesoscopic description of systems whose thermodynamic behavior is encoded in the $q$-entropy.\\ 

Before closing this section, it may be worth remembering that in the work of K.T. Sturm \cite{S1, S2} and J.Lott with C.Villani \cite{LV} in elucidating the properties of the 
Bakry-\'{E}mery Ricci curvature and in providing its synthetic definition and stability properties under measured Gromov-Hausdorff convergence, the conformal transformation of the measure
played a central role for the case of smooth metric measure spaces (Riemannian spaces with a measure). Such developments rely in a crucial way on the convexity properties of the 
$q$-entropy functional in the Wassestein space of the underlying manifold. As a result, one can see another, more indirect connection between conformal transformations on a manifold and the 
$q$-entropy. We attempted to provide a simple account of aspects of such developments in \cite{NK2}. Somewhat similar things could be said about the behavior under conformal transformations
of the Riemannian metric and volume which appreared in \cite{CGY}. One notices however that  a connection with the $q$-entropy is lacking in \cite{CGY} and does not seem likely, 
 as \cite{CGY} has a considerably different orientation and goal when compared to  the corresponding goals of \cite{S1, S2, LV}.   \\      
    

\section{Conformal changes of the metric and scalar curvature}

In some of our previous work, we also pointed out the potential significance of (quasi-)conformal maps \cite{NK1}, and of the the use of effective Riemannian metrics \ $\mathbf{g}$ \ having 
negative curvature \cite{NK4, NK5}, in describing the evolution of systems described by the $q$-entropy. Similar arguments like the ones pertinent to the conformal covariance of mesoscopic 
quantities can also be used to determine non-linear generalizations of the linear Fokker-Planck equation that are related to the $q$-entropy.\\        


\subsection{The role of the conformal Laplacian}

As was previously stated, the simplest choice would be for \ $\mathfrak{g}$ \ and for \ $\mathbf{g}$ \ to have the same form.
Given the motivation and viewpoint described above, we will focus in the sequel, in conformal changes of the metric \ $\mathbf{g}$ \ of \ $\mathcal{M}$. \ 
The time derivative in the left-hand side  the linear Fokker-Planck equation  (10) is independent of the metric, so we will concentrate on the terms of its right-hand side. 
It is well-known \cite{Sakai}  that under a conformal change of the metric 
\begin{equation}
        \widetilde{\mathbf{g}} \ = \ e^{2\phi}\mathbf{g}
\end{equation} 
the Laplacian for any sufficiently smooth function \ $f:\mathcal{M}\rightarrow\mathbb{R}$ \ becomes
\begin{equation}
           \widetilde{\nabla}^2 f \ = \ e^{-2\phi} \left\{\nabla^2 f - (n-2) ||\nabla \phi ||^2              \right\}
\end{equation}
To get a non-linear Fokker-Planck equation based on (10), one can be a minimalist and assume, in the absence of a compelling reason to the contrary, that 
\begin{equation} 
                       \phi (t, z_1,\ldots, z_k) \ = \ \phi(\rho(t_0, z_1,\ldots, z_k))
\end{equation}
where \ $\phi:\mathcal{M}\rightarrow\mathbb{R}$ \  is a function which is sufficiently smooth. We have set the evolution parameter at some particular value, for example at \  $t=0$. \ 
This is not too restrictive since our treatment relies in comparing Riemannian covariants of two metrics which are  related by a conformal transformation. 
As a result, a change in the initial value of the evolution  parameter can be absorbed into such a conformal transformation.  
The central point  here is the realization of the importance of the conformal transformations of \ $\mathbf{g}$ \ which can be seen as a predecessor, or as an outcome by hindsight, 
of the underlying  dynamics of systems described by the $q$-entropy.  \\

There is also a more concrete way to  establish non-linear Fokker-Planck equations pertinent to the $q$-entropy based on the geometry of \ $(\mathcal{M}, \mathbf{g})$. \
One way is to interpret  (10) as a gradient flow equation, following the familiar example of the heat equation, where the Dirichlet energy of the gradient flow equation is the kinetic energy. 
This viewpoint was advocated in \cite{Otto}, which was instrumental in eventually formulating synthetic notions of Ricci curvature in metric measure spaces, mainly by 
\cite{S1, S2} and \cite{LV}.             
So, the most immediate modification of (10) might be to substitute in its right-hand side instead of the usual Laplacian acting on functions of \ $\mathcal{M}$, \  the conformal Laplacian, 
which as is well-known, is given by 
\begin{equation} 
         \mathbb{L} \ =  \ 4 \ \frac{n-1}{n-2}\nabla^2 - \mathbf{R}
\end{equation}
where \ $\mathbf{R}$ \  is the scalar curvature of the metric \ $\mathbf{g}$ \ on \ $\mathcal{M}$.\  
Incidentally, we recall that the conformal Laplacian transforms covariantly under the conformal transformation (44), to be considered in the sequel,  as
\begin{equation}
  \widetilde{\mathbb{L}}f \ = \ u^{-\frac{n+2}{n-2}} \mathbb{L}(uf)
\end{equation}
for any function \ $f:\mathcal{M}\rightarrow\mathbb{R}$. \ 
Even though this proposed modification  relies heavily on the use of conformal transformations whose relevance was indicated above,
 it does not result in a non-linear Fokker-Planck equation.\\


\subsection{About scalar curvature}

Before proceeding, and because it is needed  in the sequel, one may wish to recall in this subsection a few facts about the scalar curvature \ $R$ \ of a Riemannian manifold \ $(M,g)$. \  
The scalar curvature (Ricci scalar) \cite{Sakai, Gromov2, Schoen1, Schoen2} 
of a Riemannian manifold \ $(M,g)$, \  whose metric is assumed to be of class \ $C^2$ \ at least,  is defined to be a continuous function 
\ $R: M \rightarrow \mathbb{R}$ \ satisfying the following four axioms \cite{Gromov2}: 
\begin{itemize}
     \item {\sf Additivity under Cartesian products:} \ Consider two Riemannian manifolds \ $(M_1, g_1)$ \ and \ $(M_2, g_2)$ \ with corresponding distance functions \ $d_1, \ d_2$ \ respectively, 
                          and let \ $M_1\times M_2$ \ denote their Cartesian product endowed with the Riemannian metric \ $g_1\oplus g_2$ \ which is the metric whose corresponding distance function \   
                         $d_{1\times 2}$ \ is given by 
                               \begin{equation}   
                                            d_{1\times2} = \sqrt{d_1^2 + d_2^2}
                               \end{equation}  
                         Then 
                               \begin{equation}
                                              R(M_1\times M_2, g_1\oplus g_2) \ = \ R(M_1, g_1) + R(M_2, g_2) 
                               \end{equation}
     \item {\sf Quadratic scaling:} \ Let the distance function of the Riemannian metric \ $g$ \ be indicated by \ $d_0$. \ 
                                             If \ $\lambda M, \ \ \lambda >0$ \ indicates the Riemannian manifold \ $M$ \ endowed with the distance function \ $\lambda d_0$ \ then 
                            \begin{equation}  
                                             R(\lambda M) \ = \ \frac{1}{\lambda^2} \ R(M)
                            \end{equation}
     \item {\sf Volume comparison:} \ Let \ $\mathrm{vol}$ \ indicate the Riemannian volume function, and \ $B_M(x,r)$ \ be the Riemannian ball in \ $M$ \ centered at \ $x$ \ and of radius \ $r$. \ 
                                          If the scalar 
                                                        curvatures of two $n$-dimensional manifolds \ $(M_1, g_1)$ \ and \ $(M_2, g_2)$ \ at some points \ $x_1\in M_1$ \ and \ $x_2\in M_2$ \  obey     
                            \begin{equation}                                                               
                                              R(M_1)(x_1) \ < \ R(M_2)(x_2)
                            \end{equation}
                                     then 
                            \begin{equation}
                                           \mathrm{vol} \ B_{M_1} (x_1, r) \ > \ \mathrm{vol} \ B_{M_2} (x_2, r)
                             \end{equation}
                              for all \ $r>0$ \ sufficiently small. 
       \item {\sf Normalization:} \ The unit 2-dimensional sphere of radius one \ $\mathbb{S}^2$ \ endowed with its round metric, has constant scalar curvature \ $R(\mathbb{S}^2)=2$, \ 
                            and the hyperbolic plane \ $\mathbb{H}^2$ \ has constant scalar curvature \ $R(\mathbb{H}^2) = -2$. 
\end{itemize}
Then one can prove that the function \ $R(M)$ \ exists and it is unique. It is also invariant under the isometries of the Riemannian manifold \ $(M,g)$. \ However one should be aware that 
according to \cite{Gromov2}: ``there is no single known geometric argument which would make use of such a description". This can be traced to the fact that the volume comparison axiom does not integrate to the corresponding property of balls of any radius.   \\  

For performing explicit calculations, it is well-known from local Differential Geometry \cite{Sakai} that for an $n$-dimensional  Riemannian manifold \ $(M,g)$, \ in a local coordinate system 
\  $(x^1, \ldots x^n)$, \  where the metric has components \ $g_{ij}, \ \ i,j=1,\ldots, n$, \ the connection coefficients (Christoffel symbols) are given by  
\begin{equation}
              \Gamma^k _{\  ij} \ = \  \frac{1}{2} \ g^{kl}(\partial_i g_{lj} + \partial_j g_{li} - \partial_l  g_{ij})            
\end{equation}
where \ \ $\partial_i\equiv \partial/\partial x^i, \ \  i=1,\ldots, n$ \ \  and the scalar curvature \ $R$ \ is given by
\begin{equation}
               R \ = \ g^{ij} (\partial_k \Gamma^k_{\ ij} - \partial_j \Gamma^k _{\ ik} + \Gamma^l _{\ ij} \Gamma^k _{\ kl} - \Gamma^l _{\ ik} \Gamma^k _{\ jl} )
\end{equation}
where the summation convention from \ 1 \ to \ $n$ \  is assumed for each pair of repeated indices. 
We can check that the scalar curvature is the unique, isometric invariant function on \ $M$, \ which is linear in the second derivatives of the metric \ $g$.\\ 

At this point,  we quote the well-known but non-trivial, formula for the change of the scalar curvature \ $\mathbf{R}$ \  of the metric \ $\mathbf{g}$ \  
under the conformal transformation (28). One gets 
\begin{equation}
     \widetilde{\mathbf{R}} \ = \ e^{-2\phi} \left\{ \mathbf{R} +2(n-1)\nabla^2\phi - (n-2)(n-1) \lVert\nabla\phi\rVert^2     \right\} 
\end{equation}  
where symbols with the tilde over them refer to the  metric \ $\widetilde{\mathbf{g}}$, \ and \ $\lVert\cdot\rVert$ \ indicates the norm induced by \ $\mathbf{g}$. \  
This can be rewritten for \ $n\neq 2$ \  as 
\begin{equation}
       \widetilde{\mathbf{R}} \ = \  e^{-2\phi} \left\{ \mathbf{R} + \frac{4(n-1)}{n-2} \exp \left(-\frac{n-2}{2}\phi \right) \nabla^2 \left( \exp\left( \frac{n-2}{2}\phi\right) \right)  \right\}
\end{equation}
The scalar curvature is the trace of the Ricci tensor on \ $\mathcal{M}$. \ This helps with performing local calculations in differential geometry, 
which are very important in Physics and several branches of Mathematcs,  but does not elucidate the geometric meaning of \ $\mathbf{R}$ \  which we need in order to proceed. 
To get a geometric intepretation  of the scalar curvature, we  use the comparison between volumes of balls in \ $(\mathcal{M}, \mathbf{g})$, \ and in \ $\mathbb{R}^n$ \ 
endowed with the flat metric. Let \ $B_\mathcal{M}(P,r)$ \ 
indicate a ball of radius \ $r$ \ and center \ $P\in \mathcal{M}$  \ and let \ $B_{\mathbb{R}^n}(0,r)$ \  be the corresponding ball of radius \ $r$ \ centered at the 
origin of \ $\mathbb{R}^n$. \ Then \cite{Sakai, Gromov2}
\begin{equation}  
     \frac{\mathrm{vol} \  B_\mathcal{M} (P,r)}{\mathrm{vol} \ B_{\mathbb{R}^n}(0,r)} \  = \ 1 - \frac{\mathbf{R}}{6(n+2)}r^2 + O(r^4)
\end{equation}
where \ $\mathrm{vol}$ \  stands for the Euclidean/Riemannian volume of the corresponding ball.  Therefore the scalar curvature describes deviations of volumes of small balls from those of their 
Euclidean counterparts having equal radii.   \\   

 The  metric \ $\mathbf{g}$ \ of \ $\mathcal{M}$ \ is not Euclidean in general, but Riemannian, or for velocity-dependent 
potential energies, such as in the cases of particles in electromagnetic or gauge fields, could also be considered as Finslerian, if needed. For simplicity, we confine ourselves here to the 
Riemannian case, as the case of velocity-dependent potentials can also be handled by considering connections on associated fiber bundles over Riemannian manifolds.  
Since \ $\mathcal{M}$ \ does not have to be Euclidean, none of the curvature covariant quantities constructed either by contraction, or by covariant differentiation, or by a combination of them
when applied to its Riemannian tensor  has to vanish at each point of \ $\mathcal{M}$. \  Moreover we must have in mind that, due to Liouville's theorem, we are interested  in Riemannian 
geometric quantities associated to the phase space volume. \\


\subsection{The Yamabe and Kazdan-Warner problems}

As we can see from its geometric interpretation (42),  the scalar curvature of \ $(\mathcal{M}, \mathbf{g})$ \ is a quantity related to volume deformations between a Euclidean and a Riemannian ball.
Therefore, we are essentially interested in such a mesoscopic description in determining a metric, possibly conformal to a Euclidean one, whose scalar curvature in each point \ $\mathcal{M}$ \  
is a function which needs to be determined. Ideally, such a function 
should be determined by the dynamics of the microscopic system, something that would undermine, to some extent, the present approach.
It is exactly this inability to solve the underlying dynamics explicitly that forces us to 
resort to  approximations in order to make any progress.  The above is the presrcibed curvature problem, whose formulation and solution is due, largely,
to J.L. Kazdan and F.W. Warner \cite{KW1, KW2, KW3, BE}. As a result, in order to ``derive" a non-linear 
Fokker-Planck equation, we can follow in part the Kazdan-Warner approach. \\

The Kazdan-Warner problem is the following: given a compact Riemannian manifold \ $(M, g)$, \ whose dimension will be assumed in this work to be at least 3, \  
and a function \ $f: M \rightarrow \mathbb{R}$, \ find a metric \ $\widetilde{g}$ \ conformal to \ $g$ \ whose scalar curvature is \ $f(x)$ \ for all points of \ $x\in M$, \  namely solve
\begin{equation}   
                   \widetilde{R} = f(x)
\end{equation}
The requirement of compactness of \ $M$ \ is appropriate in our case, as we actually have in mind a constant energy hypersurface of the effective phase space \ $\mathcal{M}$ \ 
as stated above. Then, if we set 
\begin{equation}
           \widetilde{g(x)} = [u(x)]^\frac{4}{p-2} g(x)
\end{equation}
where \ $u:M \rightarrow \mathbb{R}$ \ is a function \  $u>0$ \ used to encode the  conformal deformation of the initial metric \ $g$, \ 
then the Kazdan-Warner problem amounts to a solution of the equation 
\begin{equation}
            \mathbb{L}(u)  = f(x) [u(x)]^\frac{p+2}{p-2}   
\end{equation}
where\ $\mathbb{L}$ \ is the conformal Laplacian (31) and \ $p$ \ is the dimension of \ $M$. \ It should be noted, that the most straightforward case in solving such an equation is when \ $f(x)$ \ is constant over \  $M$. \ This is question posed in the Yamabe problem.\\

 The Yamabe problem \cite{Yamabe, Trudinger, Aubin, Schoen3} actually goes  further and asks, as a first step, for the 
infimum of the volume-normalized Einstein-Hilbert functional on a given conformal class of metrics and as a second step for the supremum of this quantity over the set of all conformal 
classes of metrics on \ $M$. \ 
This minimax approach sideteps the difficulty that the Einstein-Hilbert functional, which is used in the naive path integral employed in Euclidean quantum gravity approaches, for instance,
is not obviously bounded from below. In view of the above considerations, an important question may be whether the Yamabe constant of the phase space \ $\mathcal{M}$ \ 
may have any relevance for the system under study, especially for its out of equilibrium properties, aspects of which may be described by the $q$-entropy with appropriate values of \ $q$ \
\cite{Tsallis-book}.   \\


\subsection{Toward constructing a non-linear Fokker-Planck equation}

We  can  construct a non-linear Fokker-Planck equation along these lines as follows: 
we keep the evolution operator on the left-hand side of (9), (10), (13), (14) as is, as it has nothing to do with the metric \ $\mathbf{g}$ \ of the effective phase space \ $\mathcal{M}$. \ 
We assume, for the purposes of economy/minimalism of the theory that \ $u$ \ is a function of the probability density \ $\rho$, \ namely that
\begin{equation}      
              \widetilde{\mathbf{g}(x)} \ = \ \left\{ u[\rho(x)] \right\}^\frac{4}{n-2} \  \mathbf{g}(x) 
\end{equation}
and replace the ordinary Laplacian with the conformal Laplacian (31) on the right-hand side of  (9). 
Then, in the simplest case when \ $u(x) = x$, \ the non-linear Fokker-Planck equation will be
\begin{equation}  
            \frac{\partial\rho}{\partial t} = D \left( 4 \ \frac{n-1}{n-2} \ \nabla^2 \rho - \mathbf{R}(z_1, \ldots, z_n) \rho 
                                                                    - f(x)\rho^\frac{n+2}{n-2} \right)
\end{equation}
where the explicit dependence \ $\rho = \rho(t, z_1, \ldots, z_n)$ \  has been omitted for brevity, but the dependence of the 
scalar curvature \ $\mathbf{R}$ \ on the points of \ $\mathcal{M}$ \  has been made explicit. 
In (47) \ $D$ \ is assumed to be a constant as in (9), (13) which are the simplest cases.\\

  We observe that (47) is something like the parabolic analogue of the initial geometric differential equation used in the solution to the 
Yamabe and the Kazdan-Warner problems. The exponent of the non-linear term in \ $\rho$ \ has a special analytical significance:  it can be seen that 
\begin{equation}
       \frac{n+2}{n-2} \ = \ 2^\ast -1
\end{equation}
Here \ $2^\ast$ \  is Sobolev dual of \ $2$, \ which is defined by 
\begin{equation}
 \frac{1}{2^\ast} = \frac{1}{2} - \frac{1}{n}
\end{equation} 
It is also the critical power, namely \ $2^\ast$ \  is the maximal exponent \  $\eta^\ast$ \ for the Sobolev embedding 
\begin{equation}
              W^{1,\eta}(\mathcal{M}) \hookrightarrow L^{\eta^\ast}(\mathcal{M}),    \hspace{15mm}   \eta\in [1, +\infty) 
\end{equation}
to be valid. This is the classical embedding theorem of the Sobolev space \ $W^{1,\eta}$ \  into the corresponding Lebesgue space \ $L^\eta$ \ of \ $\mathcal{M}$, \ which is continuous 
but not compact \cite{Mazya}. From a physical viewpoint such an embedding is desirable, as it allows us to express solutions of
the non-linear Fokker-Planck equation in the form of familiar square integrable functions, whose physical interpretation is usually straightforward. If such an embedding were not available,
 then we might have to  resort to subtle approximation arguments for the solutions of the Fokker-Planck equation by integrable functions, a process which could obscure the direct physical 
interpretation of such solutions \ $\rho$.\\

One sees directly that even though (47) is a non-linear Fokker-Planck equation, it does not have the form of the porous medium equation (13). A suggestion in bringing these forms closer 
is to  re-parametrize  (51) by setting 
\begin{equation}    
         v(x) \ = \ [u(x)]^\frac{p+2}{p-2}
\end{equation}
then substitute the resulting expression as the right-hand side of (47) in terms of $v(\rho)$. Or, one can choose another function, instead of  $u(x)=x$ 
 in order to get an equation resembling (13).  \\
  
One can follow a different approach toward obtaining a non-linear Fokker-Planck equation by the following: 
 using the parametrization (44) of the deformation of the metric \ $\mathbf{g}$ \  of the effective phase space \ $\mathcal{M}$ \  one finds \cite{Schoen2}
\begin{equation}
          \widetilde{\mathbf{R}} \ = \ -\frac{4(n-1)}{n-2} \ u^{-\frac{n+2}{n-2}} \ \left( \nabla^2 u  - \frac{n-2}{4(n-1)}\mathbf{R}u \right)
\end{equation} 
which gives
\begin{equation}
                \nabla^2 u \ = \ -\frac{n-2}{4(n-1)}\left(\widetilde{\mathbf{R}}u^\frac{n+2}{n-2} - \mathbf{R}u  \right)
\end{equation}
Consider now the linear Fokker-Planck equation (9) and substitute (53) in its right-hand side. Then we still get a linear differential equation in \ $\rho$, \ but the deformation function 
\ $u$ \ is still undetermined. If, as in (47), we demand \ $u(x) = u(\rho(x))$ \ then we get a second order non-linear differential equation in \ $\rho$ \ due to the presence of \ 
$\widetilde{\mathbf{R}}$ \ in it.
Since the scalar curvature \ $\widetilde{\mathbf{R}}$ \ involves a non-linear combination of two derivatives of \ $\widetilde{\mathbf{g}}$ \ and their contractions as can be seen from (38), (39)
which in turn contains \ $u(\rho)$, \ one obtains a non-linear partial differential equation for \ $\rho$ \ which can be seen as the Fokker-Planck equation satisfied by \ $\rho$. \ 
The resulting differential equation is quite complicated and not particularly illuminating without further simplifying assumptions, so we will forego explicitly writing it down in its full generality. 
What exactly might some simplifying assumption be in order to make such a non-linear Fokker-Planck equation more manageable is unclear to us at this point. \\ 

We can, however, speculate that one such simplifying assumption would be  to insist that the final metric has constant scalar curvature \ $\widetilde{\mathbf{R}}$. \ 
This is a rather strong condition for the case when \ $\mathcal{M}$ \ 
has dimension 2 or 3, as it amounts to assuming that \ $\mathcal{M}$ \ has constant sectional curvature, hence that it is a sphere, a Euclidean space or a hyperbolic space. Then solution of any differential equation simplifies considerably, and the same can be said about the Fokker-Planck equations that we purport to describe. However, the 
phase spaces of interest to us  have high dimension since they correspond to systems of many degrees of freedom, even  in a mesoscopic description. It would be a rather optimistic 
to expect  the description of Hamiltonian systems on such phase spaces \ $\mathcal{M}$ \ to be effectively provided by metrics \ $\widetilde{\mathbf{g}}$ \ of constant scalar curvature.
This remains generically true,  unless we can produce an argument to the contrary, which we cannot, at this stage of development at least.   
Actually, very little if anything, is known about the behavior of such constant scalar curvature  metrics on the phase spaces \ $\mathcal{M}$ \ of mechanical systems of particles.
 Since further developing the theory  in this direction  is not important for the purposes of the present work, we will not pursue the issue any further, but defer it to future studies.  \\  
  

\section{Conclusions and discussion}

We have presented in this work a heuristic way of arriving at  non-linear Fokker-Planck equations which may provide a mesoscopic description of systems whose collective behavior 
is described by the $q$-entropy. Our arguments are hand-waving, at best. They rely crucially in the role of conformal 
transformations of the metric of the coarse grained phase space of the system \ $\mathcal{M}$. \ The role of these conformal transformations is motivated by the properties of the $q$-entropy 
and its extremizing probability distributions in pertinent variational problems, the $q$-exponentials. In such a metric approach we have pointed out the central role that is played by the scalar 
curvature of the coarse-grained metric \ $\mathbf{g}$. \\     

The motivation for the present work is to try to address aspects of the dynamical behavior  of systems conjecturally described by the $q$-entropy. Such issues are not  particularly well-understood, 
if at all. The proliferation of solutions motivated by functional forms encountered in $q$-entropy related  non-additive thermostatistics during the last twenty years,
suggests that understanding better the underlying dynamics of such systems may not only be desirable at a purely mathematical level, but aspects of them  may be analyzed in a way that may draw physically pertinent results,  with the currently available analytical techniques. \\
 
 In this work, we have not really used in any non-trivial way the fact that the underlying phase space \ $\mathfrak{M}$ \ does really have an almost complex structure $\mathbb{J}$, \  
beyond equations (2),(3),(4). Such an almost complex structure is non-integrable in general \cite{McDS}. Hence the underlying symplectic manifold is not complex, therefore cannot be K\"{a}hler 
and one has to use methods of partial differential equations to further analyze it \cite{Gromov1}. 
The original proof of the symplectic non-squeezing theorem and  the subsequent developments in the theory of pseudo-holomorphic curves are suggesting
that further progress may be made by using such techniques \cite{Gromov1, McDS2}.   However, we are not aware of any conclusions that can be drawn from implementing such pseudo-holomorphic curve techniques, of via any other method, which may be of importance to Statistical Mechanics. We suspect that the resolution of such questions may  be of some importance for a better 
understanding of the foundations, and for results pertinent to concrete models, of Statistical Mechanics.\\




\begin{thebibliography}{99}
\bibitem{vanKampen} N. van Kampen, \ \emph{Stochastic Processes in Physics and Chemistry}, 3rd Ed., \ North Holland,  \ Amsterdam, The Netherlands \ (2007). 
\bibitem{Risken} H. Risken, \ \emph{The Fokker-Planck Equation: Methods of Solution and Applications}, 2nd Ed., \ Springer-Verlag, \ Berlin, \  Germany \ (1989).
\bibitem{Frank} T.D. Frank, \ \emph{Nonlinear Fokker-Planck Equations: Fundamentals and Applications}, \ Springer-Verlag, \ Berlin, Germany \ (2005).
\bibitem{Tsallis-book} C. Tsallis, \ \emph{Introduction to Nonextensive Statistical Mechanics: Approaching a Complex World}, \ Springer Science + Business Media, \ 
                                                                        New York, NY, USA \ (2009). 
\bibitem{PP} A.R. Plastino, A. Plastino, \ \emph{Non-extensive statistical mechanics and generalized Fokker-Planck equation}, \  Physica A {\bf 222}, \ 347-354  \ (1995). 
\bibitem{TB} C. Tsallis, D.J. Bukman, \ \emph{Anomalous diffusion in the presence of external sources: Exact time-dependent solutions and their thermostatistical basis}, \ Phys. Rev. E {\bf 54}, \
                                                     R2197-R2200 \ (1996).   
\bibitem{Borland} L. Borland, \ \emph{Microscopic dynamics of the nonlinear Fokker-Planck equation: A phenomenological model}, \ Phys. Rev. E {\bf 57}, \ 6634-6642 \ (1998).
\bibitem{BPPP} L. Borland, F. Penini, A.R. Plastino, A. Plastino, \ \emph{The nonlinear Fokker-Planck equation with state dependent diffusion-a nonextensive maximum entropy approach}, 
                                                           \    Eur. Phys. J. B {\bf 12}(2), \ 285-297 \ (1998).
\bibitem{MPP} S. Martinez, A.R. Plastino, A. Plastino, \ \emph{Nonlinear Fokker-Planck equations and generalized entropies}, \ Physica A {\bf 259}, \ 183-192 \ (1998). 
\bibitem{Shiino} M. Shiino, \ \emph{Free energies based on generalized entropies and  H-theorems for nonlinear Fokker-Planck equations}, \ J. Math. Phys. {\bf 42}(6), \ 2540 \ (2001).
\bibitem{Beck} C. Beck, \ \emph{Dynamical Foundations of Nonextensive Statistical Mechanics}, \ Phys. Rev. Lett. {\bf 87}, \ 180601 \ (2001).
\bibitem{MMPL} L.C. Malacarne, R.S. Mendes, I.T. Pedron, E.K. Lenzi, \ \emph{N-dimensional nonlinear Fokker-Planck equation wih time-dependent coefficients}, \ Phys. Rev. E {\bf 65}, 
                                          \ 052101 \ (2002).
\bibitem{Frank} T.D. Frank, \ \emph{Generalized Fokker-Planck equations derived from generalized linear nonequlibrium thermodynamics}, \ Physica A {\bf 310}, \ 397-412 \ (2002).  
\bibitem{CN1} E.M.F. Curado, F.D. Nobre, \ \emph{Derivation of nonlinear Fokker-Planck equations by means of approximations to the master equation}, \ Phys. Rev. E {\bf 67}, \ 
                                                               021107 \ (2003). 
\bibitem{Chav} P.-H. Chavanis, \ \emph{Generalized Fokker-Planck equations and effective thermodynamics}, \ Physica A {\bf 340}, \ 57-65 \ (2004). 
\bibitem{NCR} F.D. Nobre, E.M.F. Curado, G. Rowlands, \ \emph{A procedure for obtaining general nonlinear Fokker-Planck equations}, \ Physica A {\bf 334}, \ 109-118 \ (2004).
\bibitem{SCN1} V. Schw\"{a}mmle, E.M.F. Curado, F.D. Nobre, \ \emph{A general nonlinear Fokker-Planck equation and its associated entropy}, \  Eur. Phys. J. B {\bf 58}, \ 159-165  \ (2007).
\bibitem{SNT} V. Schw\"{a}mmle, F.D. Nobre, C. Tsallis, \ \emph{q-Gaussians in the porous-medium equation: stability and time evolution}, \ Eur. Phys. J. B {\bf 66}, \ 537-546 \ (2008).
\bibitem{FC} M.A. Fuentes, M.O. C\'{a}ceres, \ \emph{Computing the nonlinear anomalous diffusion equation  from first principles}, \ Phys. Lett. A {\bf 372}, \ 1236-1239 \ (2008). 
\bibitem{SCN2} V. Schw\"{a}mmle, E.M.F. Curado, F.D. Nobre, \ \emph{Dynamics of normal and anomalous diffusion in nonlinear Fokker-Planck equations}, Eur. Phys. J. B {\bf 70}, \ 107-116 \ (2009).
\bibitem{ASMNC} J.S. Andrade Jr., G.F.T. da Silva, A.A. Moreira, F.D. Nobre, E.M.F. Curado, \ \emph{Thermostatistics of overdamped motion of interacting particles}, \ Phys. Rev. Lett. {\bf 105}, \
                                                              260601 \ (2010).
\bibitem{Shiino2} M. Shiino, \ \emph{Nonlinear Fokker-Planck Equations Associated with Genralized Entropies: Dynamical Characterization and Stability Analyses}, \ J. Phys. Conf. Ser. {\bf 201}, \ 
                                                                         012004 \ (2010). 
\bibitem{RNC} M.S. Ribeiro, F.D. Nobre, E.M.F. Curado, \ \emph{Classes of N-Dimensional Nonlinear Fokker-Planck Equations Associated to Tsallis Entropy}, \ Entropy {\bf 13}, \ 1928-1944 \ (2011).
\bibitem{LB} J.F. Lutsko, JP Boon, \ \emph{Microscopic theory of anomalous diffusion based on particle interactions}, \ Phys. Rev. E {\bf 88}(2), \ 022108 \ (2013).
\bibitem{ABT}  Z.G. Arenas, D.G. Barci, C. Tsallis, \ \emph{Nonlinear inhomogeneous Fokker-Planck equation within a generalized Stratonovich prescription}, \ Phys. Rev. E {\bf 90}, \ 032118 \ (2014).
\bibitem{RCN} M.S. Ribeiro, G.A. Casas, F.D. Nobre, \ \emph{Multi-diffusive nonlinear Fokker-Planck equation}, \ J. Phys A: Math. Theor.  {\bf 50}(6), \ 065001 \ (2017). 
\bibitem{SANC} A.M.C. Souza, R.F.S. Andrade, F.D. Nobre, E.M.F. Curado, \ \emph{Thermodynamic Framework for Compact q-Gaussian Distributions}, \ Physica A {\bf 491}, \ 153-166 \ (2018).
\bibitem{PCNT} A.R. Plastino, E.M.F. Curado, F.D. Nobre, C. Tsallis, \ \emph{From nonlinear Fokker-Planck equation to the Vlasov description and back: Confined interacting particles with drag}, \ 
                                                                   Phys. Rev. E {\bf 97}, \ 022120 \ (2018).
\bibitem{CN2} G.A. Casas, F.D. Nobre, \ \emph{Non-linear Fokker-Planck equations in super-diffusive and sub-diffusive regimes}, \ J. Math. Phys. {\bf 60}, \ 053301 \ (2019). 
\bibitem{NK1} N. Kalogeropoulos, \ \emph{Moduli of curve families and (quasi-)conformality of power-law entropies}, \ Int. J. Geom. Methods Mod. Phys. {\bf 13}(5), \ 1650063 \ (2016). 
\bibitem{NK2} N. Kalogeropoulos, \ \emph{Ricci Curvature, Isoperimetry and a Non-additive Entropy}, \ Entropy {\bf 17}(3), \ 1278-1308 \ (2015). 
\bibitem{McDS} D. McDuff, D. Salamon, \ \emph{Introduction to Symplectic Topology}, 3rd Ed., \ Oxford Univ. Press, \ Oxford, United Kingdom \ (2017).
\bibitem{Berger} M. Berger, \ \emph{A Panoramic View of Riemannian Geometry}, \ Springer-Verlag, \ Berlin, Germany \ (2003).
\bibitem{Gromov1} M. Gromov, \ \emph{Pseudo holomorphic curves in symplectic manifolds}, \  Invent. Math. {\bf 82}, \  307-347 \ (1985).
\bibitem{deGosson1} M. de Gosson, \ \emph{On the goodness of ``quantum blobs"  in phase space quantization}, \ {\sf arXiv:0407129 [quant-phys]}
\bibitem{deGosson2} M. A. de Gosson, \ \emph{Quamtum Blobs}, \  Found. Phys. {\bf 43}(4), \  440-457 \ (2013).
\bibitem{NK3} N. Kalogeropoulos, \ \emph{Entropies from coarse-graining: convex polytopes vs ellipsoids}, \ Entropy {\bf 17}, \  6329-6378 \ (2015).
\bibitem{Tsallis} C. Tsallis, \ \emph{Possible generalization of Boltzmann-Gibbs statistics}, \ J. Stat. Phys. {\bf 52}(1-2), \ 479-487 \ (1988).
\bibitem{Vazquez} J.L. Vazquez, \ \emph{The Porous Medium Equation: Mathematical Theory}, \ Oxford Univ. Press, \  Oxford, United Kingdom \  (2006).
\bibitem{Falconer} K. Falconer, \ \emph{Fractal Geometry: Mathematical Foundations and Applications}, 2nd Ed., \ John Wiley \& Sons Ltd., \ Chistester, United Kingdom (2003).
\bibitem{Abe} S. Abe, \ \emph{A note on the q-deformation theoretic aspect of the generalized entropies in nonextensive physics}, \ Phys. Lett. A {\bf 224}, \ 326-330 \ (1997).
\bibitem{S1} K.-T. Sturm, \ \emph{On the geometry of metric measure spaces}, \ Acta Math. {\bf 196}(1), \ 65-131 \ (2006). 
\bibitem{S2} K.-T. Sturm, \ \emph{On the geometry of metric measure spaces. II}, \ Acta Math. {\bf 196}(1),  \ 133-177 \ (2006).
\bibitem{LV} J. Lott, C. Villani, \ \emph{Ricci curvature for metric-measure spaces via optimal transport}, \ Ann. Math. {\bf 169}(3), \ 903-991 \ (2009).
\bibitem{CGY} S.-Y.A. Chang, M.J. Gursky, P. Yang, \ \emph{Conformal invariants associated to a measure}, \ Proc. Nat. Acad. Sci. {\bf 103}(8), \ 2535-2540 \ (2006).  
\bibitem{NK4} N. Kalogeropoulos, \ \emph{Tsallis entropy induced metrics and CAT(k) spaces}, \ Physica A {\bf 391}, \ 3435-3445 \ (2012).
\bibitem{NK5} N. Kalogeropoulos, \ \emph{Vanishing largest Lyapunov exponent and Tsallis entropy}, \  QScience {\bf 2013}, \ 26 \ (2013).
\bibitem{Sakai} T. Sakai, \ \emph{Riemannian Geometry}, \ Transl. Math. Monog. {\bf 149}, \ Amer. Math. Soc., \ Providence, RI, USA \ (1996). 
\bibitem{Otto} F. Otto, \ \emph{The Geometry of Dissipative Evolution Equations: The Porous Medium Equation}, \ Commun. Partial Diff. Eq. {\bf 26}(1-2),  \  101-174 \ (2001). 
\bibitem{Gromov2} M. Gromov, \ \emph{Four Lectures on Scalar Curvature}, \  29 August 2019, \ freely available online at\\
{\sf https://www.ihes.fr/~gromov/wp-content/uploads/2019/08/scalar-lectures-IHES-2019-6.pdf}
\bibitem{Schoen1} R.M. Schoen, \ \emph{Variational Theory for the Total Scalar Curvature Functional for Riemannian Metrics and Related Topics}, \
 in Topics in Calculus of Variations, \ Montecatini Terme 1987, \ pp. 120-154, \ M. Gianquinta (Ed.), \ Lectures Notes Math. {\bf 1365}, \ Springer-Verlag, Berlin, Germany  \ (1989). 
\bibitem{Schoen2} R.M. Schoen, \emph{Topics in Scalar Curvature}, Notes by Chao Li, Spring 2017, freely available online at \ \ 
{\sf http://www.homepages.ucl.ac.uk/$\sim$ucahjdl/Schoen\_Topics\_in\_scalar\_curvature\_2017.pdf}   
\bibitem{KW1} J.L. Kazdan, F.W. Warner, \ \emph{Curvature Functions for 2-manifolds with Negative Euler Characteristic}, \ Bull. Amer. Math. Soc. {\bf 78}(4), \ 570-574 \ (1972).
\bibitem{KW2} J.L. Kazdan, F.W. Warner, \ \emph{Scalar Curvature and Conformal Deformation of Riemannian Structure}, \ J. Diff. Geom. {\bf 10}(1), \ 113-134 \ (1975).
\bibitem{KW3} J.L. Kazdan, F.W. Warner, \ \emph{Existence and conformal deformation of metrics with prescribed Gaussian and scalar curvatures}, \ Ann. Math. {\bf 101}(2), \ 
                                                  317-331 \ (1975).  
\bibitem{BE} J.P. Bourguignon, J.P. Ezin, \ \emph{Scalar curvature functions in a conformal class of metrics and conformal transformations}, \ Trans. Amer. Math. Soc. {\bf 301}(2), \
                                      723-736 \ (1987).
\bibitem{Yamabe} H. Yamabe, \ \emph{On a deformation of Riemannian structures on compact manifolds}, \ Osaka Math. J. {\bf 12}(1), \ 21-37 \ (1960). 
\bibitem{Trudinger} N.S. Trudinger, \ \emph{Remarks concerning the conformal deformation of riemannian structure on compact manifolds}, \  Ann. Scuola Norm. Sup. Pisa, Serie 3,
                                {\bf 22}(2), \ 265-274 \ (1968).
\bibitem{Aubin} T. Aubin, \ \emph{\'{E}quations diff\'{e}rentielles non lin\'{e}aires et probl\`{e}me de Yamabe concernant le courbue scalaire}, \ J. Math. Pures Appl. {\bf 55}(9), 
                                                                \ 269-296 \ (1976).  
\bibitem{Schoen3} R. Schoen, \ \emph{Conformal Deformation of a Riemannian Metric to Constant Scalar Curvature}, \ J. Diff. Geom. {\bf 20}, \ 479-495 \ (1984).
\bibitem{Mazya} V. Maz'ya, \ \emph{Sobolev Spaces: with Applications to Elliptic Partial Differential Equations}, \ Grund. Math. Wissen. {\bf 342}, Springer-Verlag, \ 
Berlin, Germany (2011).
\bibitem{McDS2} D. McDuff, D. Salamon, \ \emph{J-holomorphic Curves and Symplectic Topology}, 2nd Ed., \ Colloquium Publ. {\bf 52}, \ Amer. Math. Soc., \ Providence, RI, USA \ (2012).  

\end{thebibliography}
\end{document}